# Spatio-Temporal Electromagnetic Field Shapes and their Logical Processing

G. A. Kouzaev

Department of Electronics and Telecommunications
Norwegian University of Science and Technology-NTNU
Trondheim N-7491, Norway, E-mail: guennadi.kouzaev@iet.ntnu.no

**Abstract.** This paper is on the spatio-temporal signals with the topologically modulated electromagnetic fields. The carrier of the digital information is the topological scheme composed of the separatrices-manifolds and equilibrium positions of the field. The signals and developed hardware for their processing in the space-time domain are considered.

## 1. Introduction

During last several months the increased attention has been paid to topology. Topology was established by H. Poincare, and it has been passed more then 100 years of the fruitful developments. It studies the global characteristics of objects ignoring their exact geometrical features. It is possible to find many applications of this science, but one of them is very attractive in the modeling of the intelligence. The human brain, mostly, operates with the qualitative defined information, and, the visually based processing is prevailing towards the logical handling of digital-like information. It allows defining that one of the types of human brain activity as the "spatial intelligence." The development of the logical theory and the hardware modeling is one of the most difficult tasks. The major part of the results is on the image recognizing by pixel-to-pixel methods or holographic-based optical recognition systems.

The proposed paper is on the electronic-based hardware to handle the electromagnetic, topologically modulated impulses that carry the information by their field maps and impulse magnitudes. It allows developing the electronic gates processing the images on the gate level and simulating the brain spatial activity by electronics.

## 2. Field-Force Lines Pictures and their Topological Schemes

Since the earlier years, it has been known that the fields surrounding the electrical charges and magnets have certain spatial shapes. M. Faraday used the idea of the field-force line pictures to visualize the fields and explain the found electromagnetic phenomena. Later, J.C. Maxwell used the field-force lines to create his system of equations for the electromagnetic field.

Mathematically, the field-force lines are computed by the following system of differential equations if the transient fields are pre-simulated:

$$\frac{d\mathbf{r_e}(t)}{ds_e} = \mathbf{E}(\mathbf{r}_e, t), \qquad (1)$$

$$\frac{d\mathbf{r}_h(t)}{ds_h} = \mathbf{H}(\mathbf{r}_h, t) \qquad (2)$$

where $\mathbf{r}_{e,h}$ are the radii-vectors of the field-force lines of the electric $\mathbf{E}$ and magnetic $\mathbf{H}$ fields, accordingly, $s_{e,h} = a_{e,h} t$ are the parametrical variables normalized by the coefficients $a_{e,h}$, and $t$ is the time. In general case, the equations are the non-autonomous dynamical systems, and they can be transformed to the autonomous ones by introducing a new variable $\tau$:

$$\frac{d\hat{\mathbf{r}}_\mathbf{e}}{ds_e} = \hat{\mathbf{E}}(\hat{\mathbf{r}}_e), \qquad (3)$$

$$\frac{d\hat{\mathbf{r}}_h}{ds_h} = \hat{\mathbf{H}}(\hat{\mathbf{r}}_h) \qquad (4)$$

where $\hat{\mathbf{r}}_e = (\mathbf{r}_e, \tau)$ and $\hat{\mathbf{r}}_h = (\mathbf{r}_h, \tau)$ are the radii-vectors and $\hat{\mathbf{E}} = \left( \mathbf{E}(\mathbf{r}, \tau), \frac{1}{a_e} \right)$ and $\hat{\mathbf{H}} = \left( \mathbf{H}(\mathbf{r}, \tau), \frac{1}{a_h} \right)$ are the field vectors in the extended phase-space. Then, the dimension of the phase spaces of the systems (3) and (4) is four. Besides, the vector fields $\mathbf{E}(\mathbf{r}, t)$ and $\mathbf{H}(\mathbf{r}, t)$ are governed by the Maxwell equations, boundary and initial conditions.

Qualitatively, the systems (3) and (4) are described by the ordered sets of separatrices of their extended 4-D phase spaces and the equilibrium positions where the fields are zero. These elements compose the topological scheme of a field similarly to the well-studied case on a plane [1,2]. Taking into account that the dimension of the systems (3) and (4) is four in general case, the separatrices of the phase space can be the 1-D, 2-D and 3-D manifolds. Last of them have not been studied well [3-5]. For example, only recently the Poincarè conjecture was proofed in [6,7].

In electromagnetics, the topological models of the dynamical systems (1) and (2) are used not only to analyze the pre-computed fields but to derive the qualitative solutions of the boundary electromagnetic problems [8,9]. These topological solutions of electromagnetic problem consist of analytical or semi-analytical composing of topological schemes according to the given boundary condition. It gives a rough model of the field used for qualitative estimations or as an initial approximation for further problem treatment [10].

The developed theory and derived solutions for the electromagnetic problems in the space-frequency domain allowed proposing the idea on the topological modulation of the field [8,11]. Logical variables are assigned to different topological schemes, and the signal is a series of field impulses with their discretely modulated spatio-temporal forms. In general case, these digital shapes are composed from the manifolds of different dimensions, including the 3-D ones, and the developed hardware not only detects these topologically different field shapes but processes them according to the Boolean, predicate and constraint logics [11-19].

### 3. Topologically Modulated Signals and Transmission Lines

The idea of logical handling of topological field shapes was proposed during a research on the qualitative theory of the electromagnetic boundary problems. It was found that the field topological schemes are changed discretely, and for the electric and magnetic fields, it is written each own topological scheme $T_{e,h}$, correspondingly. Taking into account the time-dependence of the fields, the considered schemes are the objects that have a certain shape in the 4-D phase space of the equations (3) and (4).

The schemes of the *i-th* and *j-th* fields can be non-homeomorphic to each other $T_{e,h}^{(i)} \not\approx T_{e,h}^{(j)}$, and then a natural number *i* or *j* can be assigned to each scheme. For example, the modes of a transmission line can have non-homeomorphic topological schemes of their fields, and they are numbered by the modal numbers. Additionally, the non-homeomorphic schemes can be composed from arbitrary combinations of modes, as well. Generally say, the number of modes and their combinations is infinite for a given transmission line, but for a certain frequency band only a few of them are the propagating modes. A set of two non-homeomorphic topological schemes of the propagating modes can correspond to the binary system $\{i = 0, j = 1\}$. Manipulation of the topological schemes according to a digital modulation signal is the topological field modulation. The digital operations with these signals are the topological computing.

To support such signals, it was considered a number of waveguides. Among them are the rectangular and corrugated waveguides, coupled strip-slot line, coupled microstrip line and coupled strip line [12]. The first ones are the dispersive waveguides, and the modal impulses excited in such waveguides have non-separable spatial and temporal field part, and they are described with the extended autonomous systems (3) and (4).

A much more studied case is the signaling along the transversal electromagnetic (TEM) waveguides like the strip transmission line. In this case, the dynamical system describing the discretely modulated signal is still non-autonomous, but each impulse of the field can be expressed by the separable spatial and temporal parts, i.e. the systems (1) and (2) are the locally autonomous ones. It allowed avoiding some difficulties to define and describe the signals and hardware for them.

An example of signal transmitted along the coupled strip line is shown in Fig. 1. The last waveguide consists of two conducting strips placed at the dielectric substrate covered by the grounded conducting plates. This line supports the even and odd modes. The first mode has the magnetic plane symmetry of the field at the central line axis. The odd mode field is symmetrical regard to the electric central plane. The modal field topological schemes of these modes are not homeomorphic to each other, and they are assigned to the logical "1" and "0", correspondingly. Then, the topologically modulated signal is a series of modal impulses as shown in Fig. 1, and the signal "topology" is changed from impulse to impulse.

Besides the field topology, the carrier of information is the impulse magnitude, and the signals introduced in [11] are the two-placed objects. The predicate logic theory for them was considered in [14].

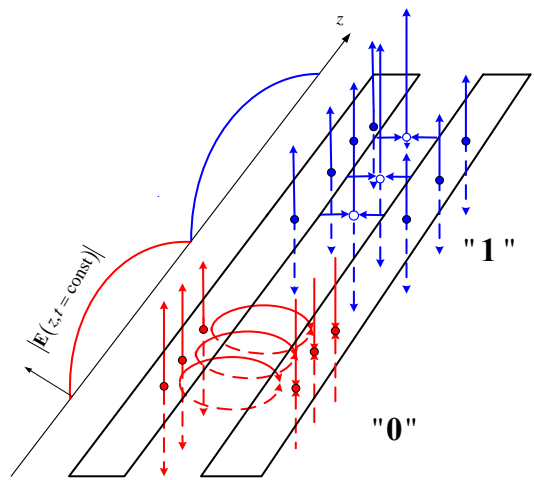

Fig. 1. Electric field of topologically modulated impulses in coupled strip transmission line at the fixed moment of time. The grounding shields are not shown.

## 4. Hardware Theory for Topologically Modulated Signals

In general, the proposed topologically modulated signals are more complicated objects then the 3-D manifolds [3-7]. The signal topological schemes consist of an ordered set of vector separatrices-manifolds, including the 3-D ones, in general case, and the field equilibrium positions.

The developed theory and circuitry allows not only detects the different spatio-temporal field shapes but compare them according to the Boolean, predicate and constrained logic. In [12-14], the reconfigurable "in-the-fly" hardware is shown. It was proposed to build such hardware using analogies with the matched filters when the outputs of a gate are tuned to a certain modal type or a spatio-temporal signal topology [12-14,15].

The theory of the components was created using the conservation law written for the power flows $p^{(n)}(\mathbf{r}_e,\mathbf{r}_h,t)$ in each *n-th* arm cross-section and the energy $W$ stored in the gate volume and expressed through the field geometry from (1) and (2) [13-16]:

$$\frac{dW(\mathbf{r}_e,\mathbf{r}_h,t)}{dt} = \sum_{n=1}^{N} p^{(n)}(\mathbf{r}_e,\mathbf{r}_h,t). \tag{5}$$

This expression reminds the Ricci flow equation used by topologists to study the manifolds [3-7]. It shows the evolution of the geometry of the excited fields to a steady state if a transient happens with the incident field. The geometry of the field force line maps inside the transformer and in the output arms depends nonlinearly on the input fields and the relationships of their amplitudes. The maps can be changed discretely by a smooth variation of the incident field parameters. Particularly, it allows controlling the field distribution of the output fields. If the output field is close to a propagating mode of the *n-th* output, then the matching conditions allow transmitting the maximal power of the incident signal to this output. Other terminals can be isolated from the input if this signal excites the evanescent modes in them. Then, the switching of the input signal from one output to another can be realized choosing the input signal parameters. Additionally, the semiconductor components can realize the time signal processing.

## 5. Logical Circuitry

The first hardware for topologically modulated signals was proposed in 1992 [11]. It deals with the signals in the waveguides with the strong dispersion to whom a complete theory (3) and (4) should be applied. As well, the low-dispersion microstrip and strip lines were considered, and the logic circuitry was patented.

The first experimental results are shown in [17] where the microwave impulses are switched by a passive circuit into different circuit arms. The developed switches allowed proposing the AND, NOT and OR gates handling the information written to the field topology.

The components for the predicate logic are considered in [14,18,19]. They allow handling the information contained in the signal amplitudes and topological schemes, and they are the two-place signals.

Additionally to the predicate circuitry, the topologically modulated impulses allow designing the reconfigurable gates [13,14]. For this purpose, one of the logical levels can be assigned to control the logical operation of the gate. Depending on the signal magnitude level, the NOT gate, for example, can be transformed into a follower for the information contained in the spatial maps of the electromagnetic signals. Next, the amplitude relationship can define the type of logical operations in a universal gate OR/AND that handling the topological charts. Then, this gate can be incorporated into a flip-flop of reconfigurable logic, and even a reconfigurable processor can be designed.

The proposed and studied two-place signals are pertinent for reversible computations. In [11], several passive circuits are studied that are reversible physically and logically. They consist of completely passive components based on the interference of the electromagnetic waves or combinations of passive and semiconductor components. Later, similar way was studied in [20].

## 6. Conclusions and Future Research

It has been considered the electromagnetic signals carrying the digital information by their spatio-temporal field structures and magnitudes. The spatio-temporal carrier is the topological scheme of the extended 4-D phase space of differential equations for field-force lines composed of the manifolds-separatrices and equilibrium field positions. The performed research touches the detecting and logical processing such topological signals. These signals are processed by the logical circuitry designed on the spatio-time filtration.

The research relates to the developments of the principles and hardware for modeling the spatial artificial intelligence [21]. Last one is defined as "a skill" to handle the spatio-temporal information. It is supposed that the information process is based on the parallel-like effects and structures, and the human mind and body handle the global spatial information, at the first, on the qualitative level [14,22,23]. It means that the topological approaches to the signal processing and computing are a key component to understand and model such a type of intellectual skills [18].

# References

1. A.A. Andronov et al, Qualitative Theory of Second Order Dynamical Systems, Transl. from Russian, Halsted Press, 1973.
2. A.A. Andronov et al, Theory of Bifurcations of Dynamical Systems on a Plane, Transl. from Russian, NASA Techn. Transl., 1971.
3. R.S. Hamilton, Three-manifolds with positive Ricci curvature, J. Differential Geometry 17 (1982), 255-306.
4. W.P. Thurston, Three dimensional manifolds, Kleinian groups and hyperbolic geometry, Bull. Amer. Math. Soc. 6 (1982) 357-381.
5. J. Milnor, Towards the Poincaré conjecture and the classification of 3-manifolds, Notices of the AMS 50 (2003), 126-1233.
6. G. Perelman, The entropy formula for the Ricci flow and its geometric applications, El. Archive: http://arxiv.org/abs/math.DG/0211159, 2002.
7. G. Perelman, Ricci flow with surgery on three-manifolds, El. Archive: http://arxiv.org/abs/math.DG/0303109, 2003.
8. G.A. Kouzaev, Mathematical fundamentals for topological electrodynamics of three-dimensional microwave integrated circuits, in: Electrodynamics and Techniques of Micro- and Millimeter Waves, pp. 37-48, Moscow, MSIEM Publ., 1991, In Russian.
9. V.I. Gvozdev and G.A. Kouzaev, Physics and the field topology of 3D-microwave circuits. Russian Microelectronics 21 (1991) 1-17.
10. G.A. Kouzaev, M.J. Deen, N.K. Nikolova, and A. Rahal, Cavity models of planar components grounded by via-holes and their experimental verification. IEEE Trans., Microwave Theory and Techniques 54 (2006) 1033-1042.
11. V.I. Gvozdev and G.A. Kouzaev, Microwave flip-flop, Russian Federation Patent, No 2054794, 02.26.1992.
12. G.A. Kouzaev, High-speed Signal Processing Circuits on the Principles of the Topological Modulation of the Electromagnetic Field, In Russian, Doctoral Thesis, Moscow, 1997.
13. G.A. Kouzaev and I.V. Nazarov, Topological impulse modulation of the fields and the hybrid logic devices, In Russian. Proc. Conf. and Exhibition on Microwave Technique and Satellite Commun. 4, Sevastopol, Ukraine, (1993) 443-446.
14. G.A. Kouzaev, Information properties of electromagnetic field superposition, J. Comun. Techn. Electronics (Radiotekhnika i Elektronika ) 40 (5) (1995) 39-47.
15. D.V. Bykov, V.I. Gvozdev, and G.A. Kouzaev, Contribution to the theory of topological modulation of the electromagnetic field, Russian Physics Doklady 38 (1993) 512-514.
16. G.A. Kouzaev, I.V. Nazarov, and V.V. Tchernyi, Circuits for ultra high-speed processing spatially modulated electromagnetic field signals, Int. J. Microcircuits and Electron. Packaging 20 (1997) 501-515.
17. V.I. Gvozdev, G.A. Kouzaev, G.M. Tchernaykov, and V.A. Shepetina, Topological demodulator, Telecommunications and Radio-Engineering 48 (1993) 26-28.
18. G.A. Kouzaev, Topological computing, Proc. 2006 WSEAS Int. Conf. Computers. Athens, Greece, July 13-15, 2006 1247-1250.
19. G.A. Kouzaev, Predicate and pseudoquantum gates for amplitude-spatially modulated electromagnetic signals, Proc. 2001 IEEE Int. Symp. Intelligent Signal Processing and Commun. Systems. Nashville, Tennessee, USA. 20-23 November, 2001.
20. B. Krishnamachari, S. Lok, C. Gracia, and S. Abraham, Ultra high speed digital processing for wireless systems using passive microwave logic, Proc. 1998 IEEE Int. Radio and Wireless Conf. (RAWCON'98), Colorado Springs, Colorado, August 1998.
21. R. Kamp, http://home.hccnet.nl/robert.kamp/.
22. G.A. Kouzaev, Theoretical aspects of the measurements of the topology of electromagnetic signals, Measurement Techniques, 39 (1996) 186-191.
23. F. Iida, R. Pfeifer, L. Steels, and Y. Kunioushi, Embodied Artificial Intelligence, Lecture Notice in Comp. Sci., Vol. 3139, Springer, 2005.